\documentclass[11pt]{article}
\usepackage{epsfig}
\usepackage{cite}

\begin{document}

\setcounter{page}{1}

\begin{center}
\begin{Large}

STAR results on longitudinal spin dynamics
\end{Large}

\vspace*{0.2in}

J. Kiryluk (for the STAR Collaboration)\\

\vspace*{0.1cm}
{\footnotesize{\em{Massachusetts Institute of Technology
\\ 77 Massachusetts Ave., Cambridge MA  02139-4307, USA\\
E-mail: joanna@lns.mit.edu
}}
} \\

\end{center}

\vspace*{0.1in}

\begin{abstract}
We present preliminary results on the double longitudinal spin asymmetries $A_{LL}$
in inclusive jet production and the longitudinal spin transfer asymmetries $D_{LL}$
in inclusive $\Lambda$ and $\bar{\Lambda}$ hyperon production.
The data amount to ~$0.5$ pb$^{-1}$ collected at RHIC in 2003 and 2004 
with beam polarizations up to $45$\%.
The jet $A_{LL}$ asymmetries, measured over $5 < p_T < 17$ GeV/c, 
are consistent with evaluations based on deep-inelastic scattering parametrizations 
for the gluon polarization in the nucleon,
and disfavor large positive values of gluon polarization in the nucleon. 
The $\Lambda$ and $\bar{\Lambda}$ $D_{LL}$, measured at midrapidity and 
at low average transverse momentum of $1.5$ GeV/c, are consistent with zero within their 
dominant statistical uncertainties.  
\end{abstract}

\vspace*{0.1in}

One of the goals of the STAR (Solenoidal Tracker At RHIC) physics program is to study the internal 
spin structure of the proton in polarized proton-proton collisions at $\sqrt{s} = 200$ 
and $\sqrt{s} = 500$~$\mathrm{GeV}$.
In particular we aim to determine the gluon polarization in the proton and the flavor 
decomposition of the quark helicity densities in the nucleon sea\cite{RHIC:Spin}.

\noindent
In this contribution we report on an exploratory measurement of the double longitudinal spin asymmetry $A_{LL}$ 
in inclusive jet production, defined as: 
\begin{equation}
A_{LL}= \frac{\sigma^{p_{+}p_{+}\rightarrow{\rm{jet}X} } - \sigma^{p_{+}p_{-}\rightarrow{\rm{jet}X}}}
{\sigma^{p_{+}p_{+}\rightarrow{\rm{jet}X}} + \sigma^{p_{+}p_{-}\rightarrow{\rm{jet}X}}}
\end{equation}
where $\sigma^{p_{+}p_{+(-)}\rightarrow{\rm{jet}X}}$ is the inclusive jet cross sections 
where the two colliding proton beams have equal(opposite) helicities.
$A_{LL}$ is sensitive to the magnitude of the gluon polarization in 
the proton for momentum fractions $0.03 < x < 0.3$\cite{Jager:2004}. In addition we present
preliminary results on $\Lambda$ and $\bar{\Lambda}$ hyperon longitudinal spin transfer asymmetries $D_{LL}$, 
defined as: 
\begin{equation}
D_{LL}=\frac{\sigma^{pp_{+}\rightarrow\Lambda_{+}X} - \sigma^{pp_{-}\rightarrow\Lambda_{+}X} }
{\sigma^{pp_{+}\rightarrow\Lambda_{+}X} + \sigma^{pp_{-}\rightarrow\Lambda_{+}X} }
\end{equation}
where $pp_{+(-)} \rightarrow \Lambda_{+} X$ is the inclusive $\Lambda$ cross section 
where only one proton beam is positively (negatively) polarized. 
$D_{LL}$ is expected to be sensitive to the polarized $\Lambda$ fragmentation functions\cite{lambda} 
and to the strange (anti-)quark polarization in the nucleon at large transverse momenta\cite{sbar}.

STAR has collected about $0.5$ pb$^{-1}$ of data at $\sqrt{s}=200$ GeV with longitudinally 
polarized beams during the initial running periods in 2003 and 2004. 
The average beam polarizations were $30-40$\%.
The Time Projection Chamber (TPC), providing tracking and particle identification, 
covered pseudorapidities $|\eta| <1.3$, and a Barrel Electromagnetic Calorimeter (BEMC) 
covered $0 < \eta < 1$ \cite{nim:2003}.
Segmented Beam Beam Counters (BBC) span $3.3 < |\eta| < 5.0$ and measure proton beam luminosity 
and transverse beam polarization components. 
Most of the data were collected with a trigger requiring coincident signals from both BBC's and 
from a BEMC tower ($\Delta \eta \! \times\! \Delta \phi \!= \! 0.05 \!\times\! 0.05$) above a transverse energy 
threshold of ~$2.5$ GeV. \\

Jets are reconstructed using a midpoint-cone algorithm\cite{CDF:2000} with a cone size 
of $0.4$ that clusters charged tracks and electromagnetic energy deposits.
Selections include the requirement of a vertex on the beam axis within $\pm 60$ cm 
of the nominal interaction point, a jet axis within a fiducial volume $0.2 < \eta_{\rm{jet}} < 0.8$,
and a TPC contribution to the reconstructed jet energy to suppress triggers caused by beam background.
The sample after selections in this analysis consists of about $3\times10^{5}$ jets 
with transverse momenta of  $5 < p_T^{\rm{jet}} < 17$ GeV/$c$. \\
Experimentally the double longitudinal spin asymmetry is defined as:
\begin{equation}
A_{LL} = \frac{1}{P_1P_2} \left[ \frac{N^{++}-R_1 N^{+-}}{ N^{++}+R_1 N^{+-} } \right],
\end{equation}

\noindent
where $N_{++(+-)}$ are the inclusive jet yields for equal (opposite) spin orientations
of the protons, $R_1=L^{++}/L^{+-}$ is the ratio of luminosities for equal and opposite proton spin orientations,
and $P_{1(2)}$ are the proton beam polarization values. 

\begin{figure}[t]
\centerline{\epsfxsize=3.0in\epsfbox{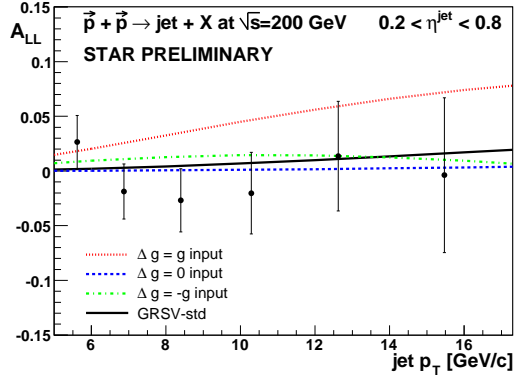}}
\caption{The longitudinal double-spin asymmetry $A_{LL}$ 
in $\vec{p}+\vec{p} \rightarrow {\rm{jet}}\;+\;X$ at $\sqrt{s}=200$ GeV versus jet $p_T$. 
The indicated uncertainties are statistical only. The curves show theoretical evaluations 
based on deep-inelastic scatetring parametrizations of gluon polarization.\label{all}}
\end{figure}

Figure~\ref{all} shows preliminary results for the double longitudinal spin asymmetry $A_{LL}$ in
inclusive jet production in polarized pp collisions at $\sqrt{s} = 200\,\mathrm{GeV}$. 
The indicated uncertainties are statistical.
We have considered systematic uncertainties from relative luminosity $R$ ($0.009$), trigger bias ($<0.007$), 
the possible contribution from residual non-longitudinal spin asymmetries ($<0.010$) and
the contamination from beam background ($0.003$). In addition, there is a scale uncertainty of $25\%$
arising from the beam polarization measurement at RHIC\cite{CNI:2003}.
Analyses with randomized spin patterns and other cross-checks including parity violating single-spin 
asymmetries show no evidence for beam bunch to bunch or fill to fill systematics.

The curves in Fig.~\ref{all} show theoretical evaluations of $A_{LL}$ in inclusive 
jet production at $\mu_F\!=\!\mu_R\!=\!p_T$ for different sets of polarized gluon distribution functions based on fits 
to deep-inelastic scattering data~\cite{Jager:2004,GRSV:2000,CTEQ6:2004}. 
They are based on a best fit to deep-inelastic scattering data (GRSV-std), and otherwise span 
the range $\Delta g(x,Q^2_0)\!=\!\pm g(x,Q^2_0\!=\!0.4\,\mathrm{GeV}^2)$~\cite{GRSV:2000} as indicated.
The data are systematically below the curve based on maximal gluon polarization, and are consistent 
with the other predictions.  Large and positive gluon polarization is thus disfavored.
 The cross section for inclusive jet production in pp collisions at 
$\sqrt{s}=200$ GeV has been measured for $5 \!< \!p_T^{\rm{jet}} \!< \!50$ GeV/$c$~\cite{Miller:2006}.
It is well described by the NLO QCD calculations\cite{Jager:2004}. \\

The $\Lambda (\bar{\Lambda})$ was reconstructed via decay modes 
$\Lambda \rightarrow p+\pi^{-} (\bar{\Lambda} \rightarrow \bar{p}+\pi^{+})$ with a branching 
ratio of $64$\%. Two tracks with opposite curvature and a topology consistent with hyperon decay were required.  
Additional cuts on the specific energy loss of protons and pions in the TPC reduced background.
The transfer asymmetry has been extracted from $\sim 30(27)$K $\Lambda (\bar{\Lambda})$ after selections, using
\begin{equation}
D_{LL} = \frac{1}{P \alpha <\cos{\theta}>} \left[ \frac{N^+ - R_2 N^{-}}{N^+ + R_2 N^{-}} \right],
\end{equation}
where 
$N_{+(-)}$ are the inclusive $\Lambda(\bar{\Lambda})$ yields for positive (negative) proton helicity
$R_2\!\!=\!\!L^{+}/L^{-}$ is the ratio of luminosities for positive (negative) proton helicity
and $P$ is the beam polarization. $\alpha\!\!=\!\!+(-)0.642 \pm 0.013$ is the empirical decay parameter for 
$\Lambda(\bar{\Lambda})$\cite{decay} 
and $\theta$ is the angle between the (anti-)proton momentum in 
the $\Lambda(\bar{\Lambda})$ rest frame and the $\Lambda(\bar{\Lambda})$ momentum direction in the laboratory frame.
$K_s$ background was suppressed by requiring $\cos\theta \!<\! -0.2$.

\begin{figure}[ht]
\centerline{\epsfxsize=3.5in\epsfbox{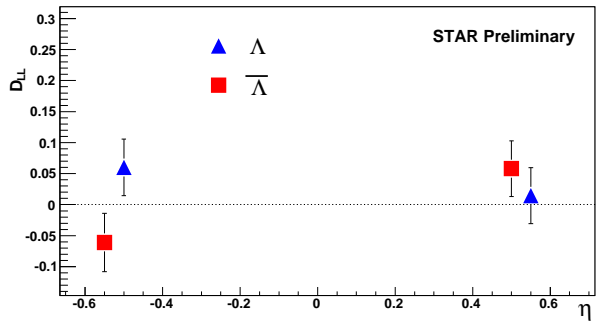}}
\caption{The longitudinal spin transfer asymmetry $D_{LL}$ 
in $p\vec{p} \rightarrow \Lambda(\bar{\Lambda})X$ at $\sqrt{s}=200$ GeV versus $\eta$.  
The indicated uncertainties are statistical only.\label{dll}}
\end{figure}

Figure 2 shows $D_{LL}$ versus $\eta$ for $\Lambda$ and $\bar{\Lambda}$. Positive $\eta$ is taken along 
the polarized beam momentum. Mean $|x_F| \simeq 8\cdot 10^{-3}$ and $p_T \simeq 1.5\,\mathrm{GeV/c}$.
The indicated uncertainties are statistical. The beam polarization measurement causes 16\% systematic 
uncertainty and the relative luminosity measurement contributes $0.01$ uncertainty.  Both are well 
below the statistical precision. The cross section for $\Lambda$ production in pp collisions at 
$\sqrt{s}=200$ GeV \cite{mark} is relatively well described by the NLO QCD calculations for $p_T>1$ GeV/c, 
which depend strongly on the choice for the parametrizations of the fragmentation functions.
Measurements of $D_{LL}$ at large transverse momenta, $p_T>8\,\mathrm{GeV/c}$, are expected to be sensitive 
mostly to the strange sea in the polarized nucleon.


\begin{thebibliography}{0}

\bibitem{RHIC:Spin}
G.~Bunce, N.~Saito, J.~Soffer, and W.~Vogelsang, Ann.~Rev.~Nucl.~Part.~Sci. {\bf{50}} (2000) 525.

\bibitem{Jager:2004}
B.~J${\rm{\ddot{a}}}$ger, M.~Stratmann and W.~Vogelsang, Phys.~Rev.~{\bf{D70}} (2004) 034010.

\bibitem{lambda}
D.~de~Florian, M.~Stratmann and W.~Vogelsang, Phys.~Rev.~Lett.~{\bf{81}} (1998)~530.

\bibitem{sbar}
Q.~Xu, Z.~Liang and E.~Sichtermann,  Phys.~Rev.~{\bf{D73}} (2006) 077503.

\bibitem{nim:2003}
Special Issue: RHIC and Its Detectors, Nucl.~Instrum.~Meth.{\bf{A499}} (2003).

\bibitem{CDF:2000}
G.~Blazey {\it et al.}, published in Batavia 1999, QCD and weak
boson physics in RunII, \\
hep-ex/0005012.

\bibitem{CNI:2003}
O.~Jinnouchi {\it et al.}, RHIC/CAD Accelerator Physics Note 171 (2004).

\bibitem{GRSV:2000}
M.~Gl${\rm{\ddot{u}}}$ck {\it et al.}, Phys.~Rev.~{\bf{D63}} (2001) 094005.

\bibitem{CTEQ6:2004}
J.~Pumplin {\it et al.}, J. High Energy Phys. {\bf{0207}} (2002) 012.

\bibitem{Miller:2006} 
M.~Miller, for the STAR Collaboration, these proceedings (2006).

\bibitem{decay} 
Particle Data Group, S.~Eidelman {\it et al.}, Phys.~Lett.~{\bf{B592}} (2004) 1. 

\bibitem{mark}
M.~T.~Heinz, for the STAR Collaboration, hep-ex/0606020. 

\end{thebibliography}
\end{document}